\documentclass[aps,achemso,floatfix,superscriptaddress,llncs]{revtex4-2}  
\usepackage{graphicx,epsfig,dcolumn,amssymb,amsmath}
\usepackage{booktabs}
\usepackage{float}
\usepackage{color}
\usepackage{lipsum}
\usepackage{hyperref}

\makeatletter
\newcommand{\printfnsymbol}[1]{%
  \textsuperscript{\@fnsymbol{2}}%
}

\begin{document}

\title{Electronic, Magnetic and Vibrational Properties of Single Layer Aluminum Oxide}

\author{A. Kutay Ozyurt}
\affiliation{Department of Physics, Izmir Institute of Technology, 35430, 
Izmir, Turkey}
 
\author{Deniz Molavali}
\affiliation{Department of Physics, Izmir Institute of Technology, 35430, 
Izmir, Turkey}

\author{Hasan Sahin}
\affiliation{Department of Photonics, Izmir Institute of Technology, 35430, 
Izmir, Turkey}

\begin{abstract}

The structural, magnetic, vibrational and electronic properties of single layer aluminum oxide (AlO$_{2}$) are investigated by performing state-of-the-art first-principles calculations. Total energy optimization and phonon calculations reveal that aluminum oxide forms a distorted octahedral structure (1T$^\prime$-AlO$_{2}$) in its single layer limit. It is also shown that surfaces of 1T$^\prime$-AlO$_{2}$ display magnetic behavior originating from the O atoms. While the ferromagnetic (FM) state is the most favorable magnetic order for 1T$^\prime$-AlO$_{2}$, transformation to a dynamically stable antiferromagnetic (AFM) state upon a slight distortion in the crystal structure is also possible. It is also shown that Raman activities (350-400 cm$^{-1}$) obtained from the vibrational spectrum can be utilized to distinguish the possible magnetic phases of the crystal structure. Electronically, both FM and the AFM phases are semiconductors with an indirect band gap and they can form a type-III vdW heterojunction with graphene-like ultra-thin materials. Moreover, it is predicted that presence of oxygen defects that inevitably occur during synthesis and production do not alter the magnetic state, even at high vacancy density. Apparently, ultra-thin 1T$^\prime$-AlO$_{2}$ with its stable crystal structure, semiconducting nature and robust magnetic state is a quite promising material for nanoscale device applications.

\end{abstract}

\maketitle

\section{Introduction}

Compounds formed by the combination of aluminum and oxygen atoms have found their use in many areas of technology. Among the possible compounds, Alumina (Al$_{2}$O$_{3}$) with its high melting point,\cite{Mavri} self-healing crystal structure,\cite{Yang} high dielectric constant\cite{Peacock} and insulating properties\cite{Vishvakarma,kim} have been widely used in electronic device applications. While bulk Al$_{2}$O$_{3}$ is a suitable material for high-throughput anti-reflective polymer lenses\cite{Yanagishita}, its thin forms have been shown to improve the spin-polarized electron tunneling in magnetic tunnel junctions.\cite{Wilt} In addition, successful synthesis of its nanowire,\cite{Meng} quantum dot\cite{Li2} and thin two-dimensional\cite{Napari,Song} forms have also been reported so far. 

Aluminum oxides can also be stabilized in the form of aluminates (Al$_{\text{x}}$O$_{\text{y}}$).  It was demonstrated that thin layers of Lanthanum-based aluminates can serve as a source of two-dimensional electron liquid exhibiting electric-field-tunable superconductivity.\cite{Cheng} It was also shown that using ultrathin Lithium aluminate nanoflakes as a polar additive, having an enormous active surface area to immobilize the lithium polysulfides, leads to extremely stable cycling performance in batteries.\cite{Ghosh} Furthermore, one-dimensional forms of strontium-europium-based aluminates are luminescent nanoribbons that can function as both light generators and waveguides, and therefore they are suitable materials for miniaturized photonic circuitry.\cite{Li} In addition to experimental studies, stabilization of single layer $\alpha$-AlO displaying indirect semiconducting behavior was also predicted theoretically.\cite{Demirci} 

At the point where nanotechnology has reached, it is necessary to further miniaturize materials, and nowadays, there is also a great need for atomically thin building blocks of all kinds. Following the first successful synthesis of graphene,\cite{novoselov2004electric} its unusual properties such as high thermal conductivity,\cite{geim2009graphene} semi-metallic behavior,\cite{Neto} and ultra-high mechanical strength\cite{lee2008measurement} have been demonstrated in a short period of time. Studies in the following years have revealed that ultra-thin stable phases of other graphene-like materials may also exist. Among them especially h-BN with its insulating properties,\cite{Li3} MoS$_2$ with its thickness-dependent electronic and optical properties,\cite{Mak,wang2012electronics} ReS$_{2}$ for its in-plane anisotropic and thickness-independent properties\cite{tongay2014monolayer} and silicene with its ultra-high carrier mobility\cite{silicene2009} have been the most prominent ones.

Although many graphene-like materials have been successfully synthesized to date, very few of them show magnetic properties. Recently, it has been found that the itinerant ferromagnetism persists in 2D Fe$_{3}$GeTe$_{2}$ down to the monolayer with an out-of-plane magnetocrystalline anisotropy.\cite{Deng} In addition, intrinsic ferromagnetic half-metallic behavior of monolayer square CrBr$_{2}$ and the strain-dependent phase transitions were predicted using first-principles calculations.\cite{Li4} Very recently, growth of iron chloride (FeCl$_{2}$) films on Au(111) and graphite with atomic thickness by molecular-beam epitaxy (MBE) technique has also been reported.\cite{zhou2020atomically} Besides, we showed that the magnetic ground state of single layer FeCl$_{2}$ crystal can be modulated by the point defects and structural stability was maintained despite the heavy defect doping.\cite{ceyhan2021}

In this work, we investigated the ultra-thin graphene-like crystal structure of aluminum oxide (AlO$_{2}$) by means of ab-initio calculations. The rest of the paper is organized as follows; computational methodology is given in Sec. \ref{section:computational}. Possible magnetic phases and their vibrational characteristics are investigated in Sec. \ref{sec:resultsA}. In addition, electronic properties of AFM and FM phases and their possible vdW type heterostructures are investigated in Sec.\ref{sec:resultsB}. Formation of single O vacancy and resulting modifications in electronic properties are given in Sec.\ref{sec:resultsC}. Finally, the conclusions are presented in Sec.\ref{sec:conclusions}.

\begin{table*}
\centering
\small
  \caption{For the different magnetic phases of 1T$^\prime$-AlO$_{2}$: lattice constants for rectangular unit cell (a and b), thickness ($h$), the amount of donated (received) electron ($\rho{}_\text{Al}$ and $\rho{}_\text{O}$), the work function ($\phi$), calculated cohesive energy for per atom (E$_{Coh}$), the band gap value (E$_{Gap}$), total magnetic moment for per the formula, ($\mu$), in-plane stiffness along zigzag direction, ($C_{zz}$), in-plane stiffness along armchair direction, ($C_{arm}$), Poisson ratio along zigzag direction, ($\upsilon_{zz}$), Poisson ratio along armchair direction, ($\upsilon_{arm}$), calculated hole effective mass (m$_h^*$), calculated electron effective mass (m$_e^*$) where m$_0$ is the free electron mass.}
  \begin{tabular*}{1\textwidth}{@{\extracolsep{\fill}}lcccccccccccccccccc}
    \hline
  	\hline
   &   & a & b & h  & $\rho{}_\text{Al}$  & $\rho{}_\text{O}$ &  $\phi$ & E$_{Coh}$ & E$_{Gap}$ & $\mu$ & $C_{zz}$ & $C_{arm}$ & $\upsilon_{zz}$  & $\upsilon_{arm}$ & m$_h^*$ & m$_e^*$ \\
    &   &  (\AA) & (\AA) & (\AA)& (e$^\text{-}$) &  (e$^\text{-}$) &  (eV) & (eV/atom) &  (eV) & ($\mu_B$) & (N/m) & (N/m) & - & - & (m$_0$) & (m$_0$) \\
    \hline \\
    & FM    & 3.05 & 4.84 & 1.77  & 0.54  & 7.23 &  8.01 & 5.57 & 0.63 & 1  &  126 & 138 &  0.40 &  0.36 & 2.29 & 1.33 & \\ 
    & AFM  & 3.06 & 4.82 & 1.78  & 0.54  & 7.23 &  7.96 & 5.56 & 1.18 & 0  &  108 & 123 &  0.33 &  0.37 & 1.20/18.34 & 8.52 &\\
	\hline
    \hline
  \end{tabular*}
\end{table*}

\begin{figure}
\centering
 \includegraphics[width=8.5cm]{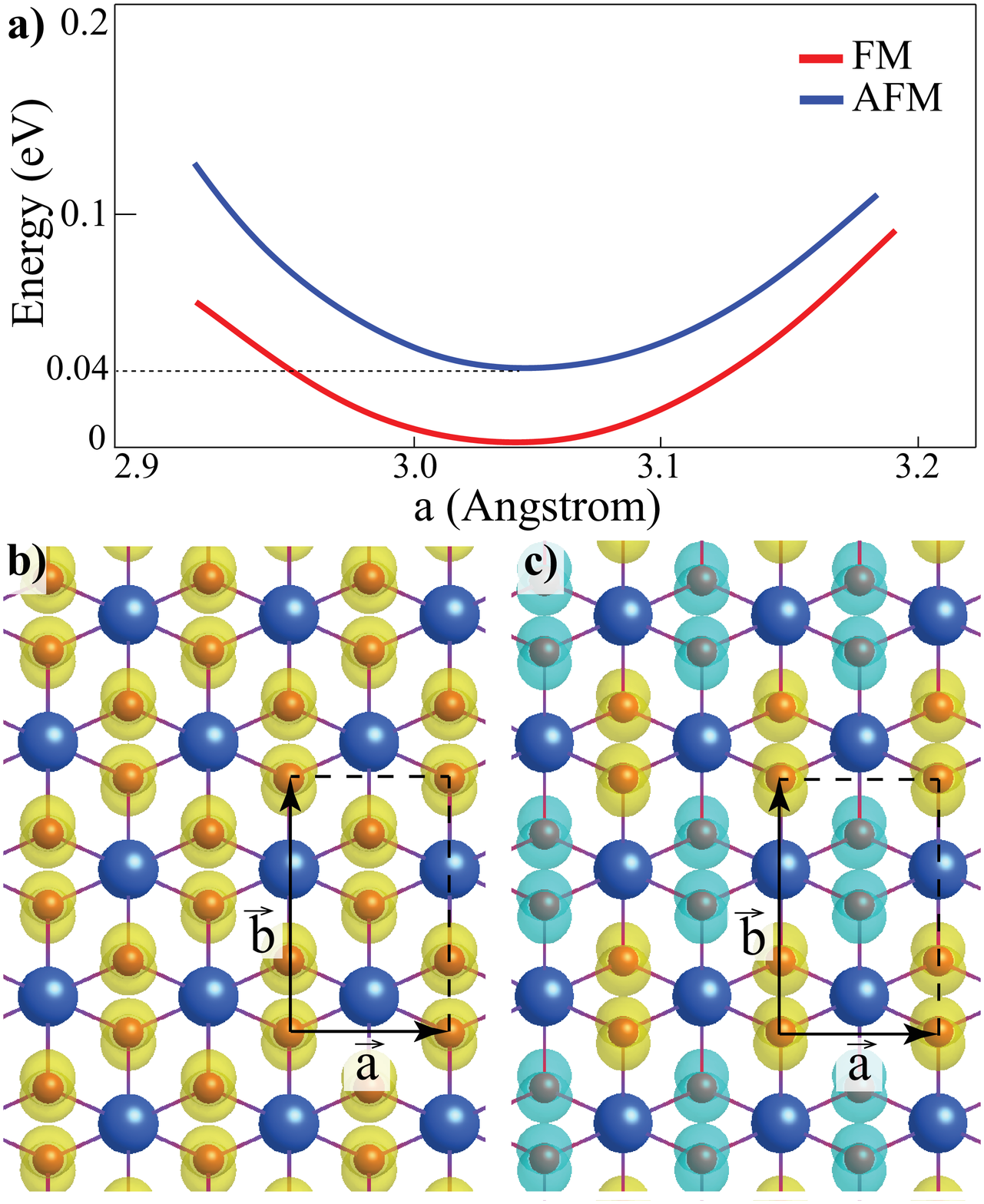}
  \caption{(a) Energy versus lattice constant (a) of 2D 1T$^\prime$-AlO$_{2}$. Top views of the geometric structures of (b) FM and (c) AFM state where blue and red atoms stand for Al and O, respectively. For 3D plot of magnetization charge densities minority and majority states are shown by blue and yellow color.}
  \label{fig:1}
\end{figure}

\section{Computational Methodology}\label{section:computational}

For determination of optimized crystal structures and resulting electronic and magnetic characteristics of AlO$_{2}$ first-principles calculations were performed by using Vienna ab initio Simulation Package (VASP).\cite{Grimme,Kresse,Kresse2} The exchange-correlation potential was treated using the Perdew-Burke-Ernzerhof Generalized Gradient Approximation (PBE-GGA).\cite{Perdew} The projector augmented wave (PAW) potentials were used as pseudo-potentials.\cite{Blochl} Van der Waals (vdW) correction within the form of DFT-D2 method of Grimme is used for approximation of dispersion forces.\cite{Grimme} Bader charge technique was used for determination of charge for each atom in the crystal structure.\cite{Henkelman}

The kinetic energy cutoff and the convergence criteria for the total energy were set to 500 eV and 10$^{-5}$ eV, respectively. At least 18.2 {\AA} vacuum spacing was inserted in order to avoid adjacent layer-layer interactions along the vertical direction. For the structural optimization of the primitive unit cell, k-space was sampled by using 9$\times$9$\times$1 k-point mesh, and doubled to 18$\times$18$\times$1 in density of states calculations. Structural optimizations were carried on until the pressure is in between $\pm$1 kB along the each lattice direction. The cohesive energy per atom was calculated using the equation $E_{Coh}=\frac{1}{n_{tot}}[n_{Al}E_{Al}+n_OE_O-E_{SL}]$ where $n_{Al}$ ($n_{O}$) is the number of Al (O) atoms in the system, while $E_{Al}$ and $E_{O}$ refer to individual energies of Al and O atoms, respectively. $E_{SL}$ is the total energy of the corresponding single layer and $n_{tot}$ is the total number of atoms per unit cell.

\begin{figure*}
  \includegraphics[width=18cm]{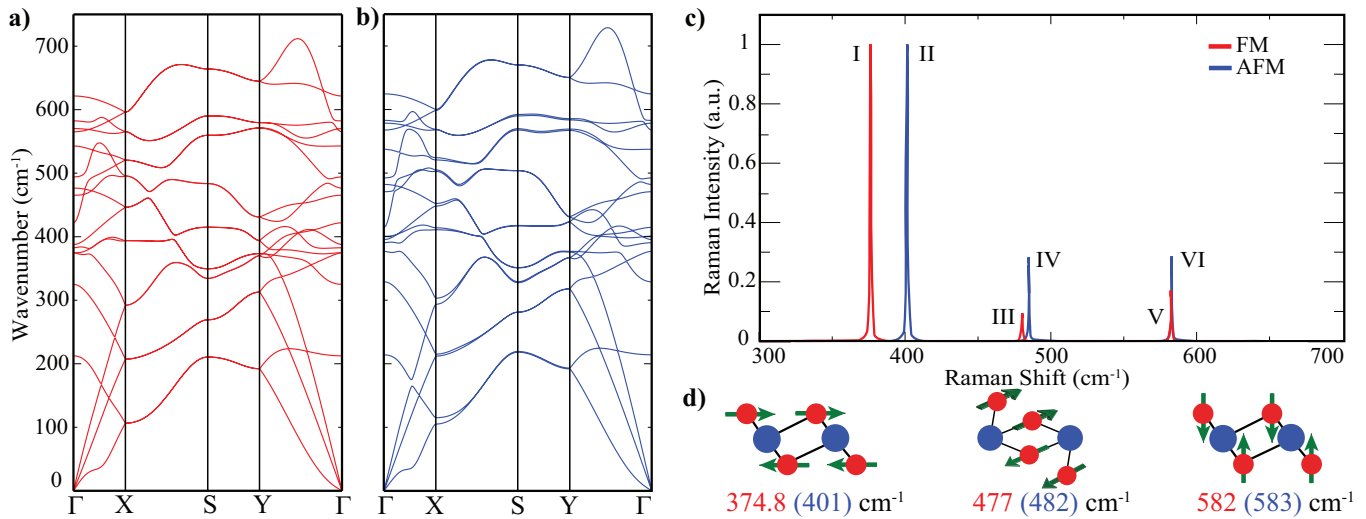}  
\caption{Phonon band structure of 1T$^\prime$-AlO$_{2}$ calculated for (a) the FM and (b) the AFM states. (c) The Raman activity spectrum of the zone-centered phonon modes. (d) The representative images of Raman active phonon vibrations.}
  \label{fig:2}
\end{figure*}

The vibrational properties were investigated via PHONOPY code, which uses the small displacement method to generate the force constant matrix.\cite{Togo} In addition, the zone-centered phonon frequencies and the corresponding off-resonant Raman activities were calculated using the finite-difference method. Basically, in a Raman experiment instantly scattered photons, whose dispersion with respect to a shift in frequency gives the Raman spectrum, are collected. The treatment of Raman activities of each vibration is based on Placzek’s classical theory of polarizability in which the activity of a Raman active mode is proportional to the square of the term, $|\hat{e}_sR\hat{e}_i|$ where $\hat{e}_s$ and $\hat{e}_i$ stand for the polarization vectors of the scattered and the incident light, respectively. In addition, $R$ represents the Raman tensor whose elements are the derivatives of the polarizability of the material with respect to the phonon normal modes. Notably, for a Raman active phonon mode some of the elements of Raman tensor are non-zero.

\section{Results}

\subsection{Structural, Magnetic and Vibrational Properties} \label{sec:resultsA}

Characteristic vibrational, electronic and magnetic fingerprints of any material are directly determined by its crystal symmetry. Therefore, a reliable prediction of the ground state crystal structure is of importance. Total energy optimization calculations show that (see Fig.\ref{fig:1}(a))  the AlO$_{2}$ crystal has a six-atom rectangular unit cell with the lattice parameter of \textbf{a} (\textbf{b}) is 3.05 (4.84) {\AA}. Here the optimized structure of AlO$_{2}$ displays slightly distorted 1T hexagonal lattice (namely 1T$^\prime$-AlO$_{2}$), where the central Al subplane is surrounded by an octahedral type arrangement of O atoms. It worths mentioning that 1T$^\prime$-AlO$_{2}$ displays ferromagnetic (FM) ground state with 1 $\mu_B$ per formula. Magnetic charge density distribution ($\rho_{\uparrow}$-$\rho_{\downarrow}$) shown in Fig. \ref{fig:1}(b) reveals that the ferromagnetic behavior stems from \textit{p$_{z}$} orbitals of O atoms. However, there is also a possible antiferromagnetic (AFM) phase of the crystal structure which is $\sim$40 meV less preferable than the FM one. Here cohesive energy of FM and AFM phase is calculated to be 5.57 and 5.56 $eV/atom$, respectively. In the AFM phase, that emerges as a result of the neighboring oxygens approaching each other, ferromagnetically ordered AlO$_{2}$ chains are formed and adjacent chains prefer to be antiferromagnetically ordered with each other. Compared to the FM phase, as a result of closer packing of AlO$_{2}$ chains in the AFM phase, there is a narrowing in the \textbf{b} lattice parameter and a slight increase in the layer thickness (defined as the distance between the neighboring subplanes of O atoms). 

Bader charge analysis shows that crystal lattice is constructed with Al-O bonds having ionic character, where the charge donation from Al to each O is 1.23 $e^\text{-}$ for both FM and AFM states.  Additionally, the work functions of the FM and AFM phases are calculated to be 8.01 and 7.96 eV, respectively, which are significantly higher than that of single layer structures of MoS$_{2}$ (6.11 eV), WS$_{2}$ (5.89 eV), MoSe$_{2}$ (5.49 eV) and h-BN (4.7 eV).\cite{Lanzillo,Pierucci} Apparently, such materials having high work function are desirable for their use as charge extraction and injection layers in electronic and optical devices in order to improve device capability.\cite{Bivour,Schulz}  

For a discussion on resistance of the 2D crystal structure distortion under mechanical load, determination of in-plane stiffness ($C$) and the Poisson ratio ($\upsilon$) is of importance. Due to its anisotropic nature, the 1T$^\prime$ phase of AlO$_{2}$ possesses anisotropy by means of its elastic constants. Here, the two elastic constants along with the zigzag ($zz$) and armchair ($arm$) directions are calculated by using the equations\cite{ma2018two} $C_{zz}=\frac{1}{C_{22}}[C_{11}C_{22}-C_{12}^{2}]$ and $C_{arm}=\frac{1}{C_{11}}[C_{11}C_{22}-C_{12}^{2}]$. Accordingly, in-plane stiffness values of FM and AFM phases are found to be 126(138) and 108(123) N/m for FM and AFM phases along the $zz$($arm$) directions, respectively. Compared with the single layer Al$_{2}$O$_{3}$ (48 N/m)\cite{Song}, AlO$_{2}$ seems to be a much stiffer material. On the other hand, as compared to single layer of AlN(116 N/m)\cite{sahin35}, AlO$_{2}$ phases are found to possess similar mechanical features in terms of the in-plane stiffness.

The Poisson ratio, which is defined as the ratio of the transverse contraction strain to the longitudinal extension, describes the mechanical response of a material to the external load. The Poisson ratio through the $zz$ and $arm$ directions are calculated by using the equations\cite{ma2018two} $\upsilon_{zz}=\frac{C_{12}}{C_{22}}$ and $\upsilon_{arm}=\frac{C_{12}}{C_{11}}$. It is known that for a stable linear elastic material, the Poisson ratio varies between -1.0 and 0.5.\cite{Gercek} Here, the Poisson ratios  of FM (AFM) phases of 1T$^\prime$-AlO$_{2}$ are calculated to be 0.40 (0.33) along $zz$ direction and 0.36 (0.37) along $arm$ direction. These values indicate the soft nature of AlO$_{2}$ when compared with those of the well-known 2D structures, such as graphene and MoS$_2$, for which the Poisson ratios are reported to be 0.19 and 0.26, respectively.\cite{lee2008measurement,Yagmurcukardes}

The dynamical stability of 1T$^\prime$-AlO$_{2}$ in FM and AFM states is also examined through their phonon band dispersions, which are obtained by utilizing the small displacement method.  Phonon dispersions depicted in Fig.\ref{fig:2} show that all the phonon branches have real eigenfrequencies through the whole Brillouin zone and therefore both FM and AFM phases of 1T$^\prime$-AlO$_{2}$ crystal are dynamically stable. It is seen that distorted crystal structure of 6-atom primitive unit cell of the crystal structure leads to presence of 18 phonon branches. The highest frequency optical phonon branch is found at 622 (625) \text{cm$^{-1}$} for the FM (AFM) order which is lower than those single layer AlN ($\sim$ 900\text{cm$^{-1}$}), Al$_{2}$O$_{3}$ (1500\text{cm$^{-1}$}), and $\alpha$-AlO ($\sim$ 650\text{cm$^{-1}$}).\cite{sahin35,Song,Demirci}

For further discussion of vibrational characteristics, zone-centered first-order off-resonant Raman spectra of FM and AFM phases are also calculated. As shown in Fig. \ref{fig:2}(c-d), The Raman active modes labeled by numbers I-II, which have the frequencies of 374.8 (401) cm$^{-1}$ for FM (AFM) states, are attributed to the shear-like vibration of upper and lower O atoms towards the Al atoms. In addition, the phonon modes III-IV are calculated to be at the frequencies 477 (482) cm$^{-1}$ for FM (AFM) states and are shown to arise from the shear-like vibration of O atoms parallel to the Al atoms. Moreover, the Raman active modes, V-VI having frequencies 582 (583) cm$^{-1}$ for FM (AFM), represent the out-of-plane vibration of upper and lower O atoms with respect to each other. While the two magnetic phases are distinguishable in terms of the frequencies of the Raman active modes, the calculated activities reveal their distinct features. The Raman activities of the most prominent modes (I-II) are found to be the same while the activities of III and V in the FM phase are found to be lower as compared to that of IV and VI in the AFM state.

\begin{figure}
\centering
  \includegraphics[width=12cm]{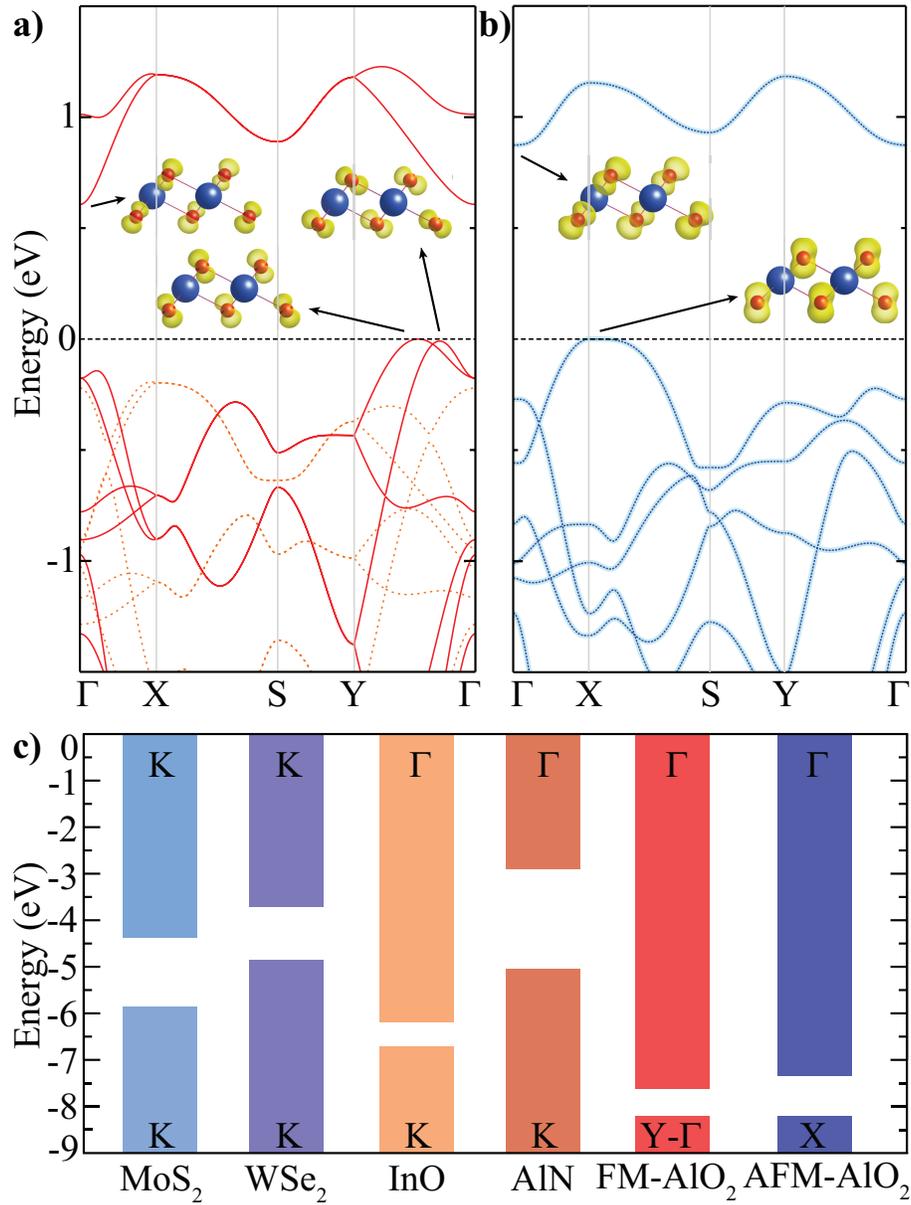}
  \caption{Electronic band structure of 1T$^\prime$-AlO$_{2}$ for (a) FM and (b) AFM phase. Dashed and solid lines show the majority and minority spin states which are degenerate in AFM state. (c) Comparative band alignments of 1T$^\prime$-AlO$_{2}$ phases with MoS$_{2}$, WSe$_{2}$, InO and AlN single layers.}
  \label{fig:3}
\end{figure}

\subsection{Electronic Properties}\label{sec:resultsB}

Electronic characteristics of FM and AFM phases of single layer 1T$^\prime$-AlO$_{2}$ are investigated through their electronic band dispersions and the band decomposed charge densities. As shown in Fig. \ref{fig:3}(a), FM phase of AlO$_{2}$ is a ferromagnetic semiconductor with an indirect bandgap of 0.63 eV. As a consequence of the large splitting in majority and minority states valence band maximum (VBM) and conduction band minimum (CBM) of the FM phase is composed of the same spin type. While the edge of the VBM reside at the Y-$\Gamma$ point, the CBM is located at the Brillouin zone center. It is also seen that both VBM and CBM band edge states have the same tilted O-$p_z$ character. As shown in Table I, the hole effective mass at the vicinity of VBM is found to be 2.29 m$_0$,  while the electron effective mass at the CBM is calculated to be 1.33 m$_0$ (where m$_0$ is electron mass). 

In addition, AFM phase displays indirect semiconducting behavior with 1.18 eV bandgap where the VBM and CBM are located at the X and the $\Gamma$ of the BZ, respectively. It appears that VBM states stem from orbitals perpendicularly oriented to the crystal surface leads to the formation of quite flat bands around the X point and therefore relatively large hole effective mass for the AFM state. Moreover, the hole effective mass is found to be anisotropic and two different values are calculated along with the $\Gamma$-X and X-S directions (1.20 and 18.34 m$_0$, respectively) while the electron effective mass at the vicinity of the CBM is found to be 8.52 m$_0$. Apparently, lattice contraction originating from the formation of AlO$_{2}$ chains leads to the much heavier charge carriers in the AFM phase. 

Considering a nano device in which a 2D ultra-thin material is used as the building block, it is clear that within the device architecture the material will form interfaces with other materials that can be thought as heterostructures. Here, we also investigate the vertical vdW heterostructures of 1T$^\prime$-AlO$_{2}$ phases constructed with the 2D crystals such as MoS$_2 $, WSe$_2$, InO, and AlN. As presented in Fig. \ref{fig:3}(c), as a consequence of their high photoelectric threshold, energies of VBM and the CBM of both phases of 1T$^\prime$AlO$_{2}$ are much lower than that of the other materials considered. Therefore, it can deduced that either FM- or AFM-phase of 1T$^\prime$-AlO$_{2}$ form type-III heterojunction with single layers of MoS$_2 $, WSe$_2$, InO, and AlN. Obviously, such type-III vdW heterostructures might exhibit a negative differential resistance phenomenon due to the existence of band-to-band tunneling as the electrons have a tendency to tunnel from VBM of these crystal structures to the CBM of 1T$^\prime$AlO$_{2}$.\cite{Ionescu,Lu,BTT}

\begin{figure}
\centering
  \includegraphics[width=10cm]{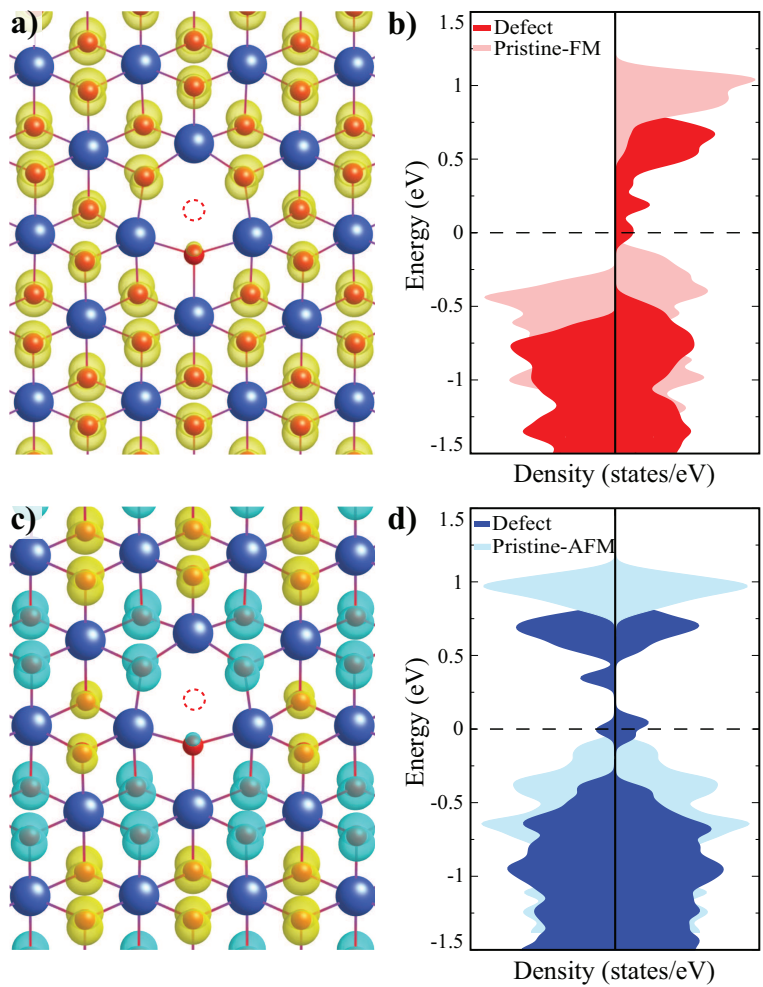}
  \caption{The atomic structure, magnetization charge density and corresponding electronic density of states of single oxygen defected 1T$^\prime$-AlO$_{2}$ in (a-b) FM and (c-d) AFM phase. The missing atom is shown by dashed red circle. }
  \label{fig:4}
\end{figure}

\subsection{Effect of Oxygen Vacancies}\label{sec:resultsC}

Whether the synthesis technique is a top-down or bottom-up approach, the formation of point defects in two-dimensional crystals is inevitable. In order to investigate the effect of atomic vacancies on the electronic properties of single layer 1T$^\prime$-AlO$_{2}$, the formation of a single O vacancy, \text{V$_{O}$}, is considered and the defected crystal structure is fully optimized. In order to simulate the single O vacancy, a supercell structure containing 96 atoms is considered and an O atom is removed from the layer. After full optimization of the O-vacant single layer of AlO$_{2}$, it is shown that structural deformations around the defect site are compensated by the local distortion of three Al atoms which tend to form Al-dimers as the hole gets larger via O vacancy. While the Al-Al distance is 2.84 {\AA} in the host lattice, that of between the Al atoms around the defect site is 2.75 {\AA} as a result of the structural distortion. On the other hand, by the formation of the vacancy, O atoms around the defect site move towards the Al atoms surrounding the defect site and thus, the Al-O bond lengths get smaller. It is also found that the formation of \text{V$_{O}$} shrinks the lattice by 0.4\% along the $x$-direction while it locally enlarges the host lattice by 0.3\%. 

The defect formation energy is calculated using the relation; $E_{f}=[E_{tot}\text{(def)}+E_{\text{O}}]-E_{tot}\text{(pristine)}$ where E$_{tot}$\text{(def)} and E$_{tot}$ represents the total energies of the defected and the pristine single layers, respectively while E$_O$ stands for the single O energy. Here, the calculated defect formation energy of \text{V$_{O}$} is 6.24 eV which indicates that it is very difficult to subsequently create an O vacancy in the single layer 1T$^\prime$-AlO$_{2}$. 

Moreover, magnetic ground state calculations reveal that FM interaction is still energetically more favorable in defected 1T$^\prime$-AlO$_{2}$. It is found that the formation of O vacancy leads to a decrease in the total magnetic moment of the structure (2$\mu_B$). Apparently, the creation of a single O vacancy induces local magnetic moments which are altered by the deviations from the nominal stoichiometry. The effect of single O vacancy on the electronic properties of 1T$^\prime$-AlO$_{2}$ is investigated in terms of the total electronic density of states. Apparently, O vacancy induces midgap states that result in a transition to the metallic state in the oxygen vacant domains of the 1T$^\prime$-AlO$_{2}$ crystal.
As shown in Fig. \ref{fig:4}(c), similar to the FM phase, the presence of the oxygen vacancy does not cause a remarkable change in the magnetic ground state when the crystal structure is in the AFM phase. It is also seen that while the magnetic order of the AFM remains unchanged, a magnetic moment of 2$\mu_B$ per oxygen occurs because the balance between the sublattice atoms is disturbed due to the missing oxygen atom. Additionally, as shown Fig. \ref{fig:4}(d) formation of oxygen vacancies also leads to a semiconductor-to-metal transition at the vicinity of defect domains.

\section{Conclusions}\label{sec:conclusions}
In conclusion, the first-principles calculations were performed on single layer AlO$_{2}$ and its structural, magnetic, vibrational and the electronic properties were investigated. Total energy optimizations revealed that single layer 1T$^\prime$-AlO$_{2}$ possesses two distinct magnetic phases, namely FM and AFM. While the FM phase of 1T$^\prime$-AlO$_{2}$ was found to be energetically more favorable over the AFM one, both phases are dynamically stable and they are distinguishable in terms of their Raman spectra. Electronically, both FM- and AFM-1T$^\prime$-AlO$_{2}$ structures were shown to be semiconductors with an indirect band gap. It is also seen that 1T$^\prime$-AlO$_{2}$ crystals form a broken-gap type-III heterojunction when combined with other graphene-like ultra-thin materials. Moreover, we predicted that presence of oxygen vacancies yields semiconductor-to-metal transition without altering the magnetic order in the crystal structure. The ultra-thin 1T$^\prime$-AlO$_{2}$ with its semiconducting nature and robust magnetic ground state unaffected by the presence of structural defects is a promising material for nanoscale device applications.

\section{Acknowledgments}
Computational resources were provided by TUBITAK ULAKBIM, High Performance and Grid Computing Center (TR-Grid e-Infrastructure).

\bibliography{references}

\providecommand{\latin}[1]{#1}
\makeatletter
\providecommand{\doi}
  {\begingroup\let\do\@makeother\dospecials
  \catcode`\{=1 \catcode`\}=2 \doi@aux}
\providecommand{\doi@aux}[1]{\endgroup\texttt{#1}}
\makeatother
\providecommand*\mcitethebibliography{\thebibliography}
\csname @ifundefined\endcsname{endmcitethebibliography}
  {\let\endmcitethebibliography\endthebibliography}{}
\begin{mcitethebibliography}{46}
\providecommand*\natexlab[1]{#1}
\providecommand*\mciteSetBstSublistMode[1]{}
\providecommand*\mciteSetBstMaxWidthForm[2]{}
\providecommand*\mciteBstWouldAddEndPuncttrue
  {\def\EndOfBibitem{\unskip.}}
\providecommand*\mciteBstWouldAddEndPunctfalse
  {\let\EndOfBibitem\relax}
\providecommand*\mciteSetBstMidEndSepPunct[3]{}
\providecommand*\mciteSetBstSublistLabelBeginEnd[3]{}
\providecommand*\EndOfBibitem{}
\mciteSetBstSublistMode{f}
\mciteSetBstMaxWidthForm{subitem}{(\alph{mcitesubitemcount})}
\mciteSetBstSublistLabelBeginEnd
  {\mcitemaxwidthsubitemform\space}
  {\relax}
  {\relax}

\bibitem[Mavri{\v{c}} \latin{et~al.}(2018)Mavri{\v{c}}, Fanetti, Mali, and
  Valant]{Mavri}
Mavri{\v{c}},~A.; Fanetti,~M.; Mali,~G.; Valant,~M. High-temperature
  stabilization of bulk amorphous Al$_2$O$_3$. \emph{J. Non-Cryst. Solids}
  \textbf{2018}, \emph{499}, 363--370\relax
\mciteBstWouldAddEndPuncttrue
\mciteSetBstMidEndSepPunct{\mcitedefaultmidpunct}
{\mcitedefaultendpunct}{\mcitedefaultseppunct}\relax
\EndOfBibitem
\bibitem[Yang \latin{et~al.}(2018)Yang, Kushima, Han, Xin, and Li]{Yang}
Yang,~Y.; Kushima,~A.; Han,~W.; Xin,~H.; Li,~J. Liquid-like, self-healing
  aluminum oxide during deformation at room temperature. \emph{Nano Lett.}
  \textbf{2018}, \emph{18}, 2492--2497\relax
\mciteBstWouldAddEndPuncttrue
\mciteSetBstMidEndSepPunct{\mcitedefaultmidpunct}
{\mcitedefaultendpunct}{\mcitedefaultseppunct}\relax
\EndOfBibitem
\bibitem[Peacock and Robertson(2002)Peacock, and Robertson]{Peacock}
Peacock,~P.; Robertson,~J. Band offsets and Schottky barrier heights of high
  dielectric constant oxides. \emph{J. Appl. Phys.} \textbf{2002}, \emph{92},
  4712--4721\relax
\mciteBstWouldAddEndPuncttrue
\mciteSetBstMidEndSepPunct{\mcitedefaultmidpunct}
{\mcitedefaultendpunct}{\mcitedefaultseppunct}\relax
\EndOfBibitem
\bibitem[Vishvakarma \latin{et~al.}(2021)Vishvakarma, Kannan, Vashishtha, and
  Pal]{Vishvakarma}
Vishvakarma,~S.; Kannan,~R.; Vashishtha,~N.; Pal,~A.~S. Evaluation of hybrid
  sol-gel alumina coating for electrical resistance applications. \emph{Surf.
  Topogr.} \textbf{2021}, \emph{9}, 045030\relax
\mciteBstWouldAddEndPuncttrue
\mciteSetBstMidEndSepPunct{\mcitedefaultmidpunct}
{\mcitedefaultendpunct}{\mcitedefaultseppunct}\relax
\EndOfBibitem
\bibitem[Kim \latin{et~al.}(2016)Kim, Ha, Lee, and Song]{kim}
Kim,~J.; Ha,~J.-H.; Lee,~J.; Song,~I.-H. Optimization for permeability and
  electrical resistance of porous alumina-based ceramics. \emph{J. Korean
  Ceram. Soc.} \textbf{2016}, \emph{53}, 548--556\relax
\mciteBstWouldAddEndPuncttrue
\mciteSetBstMidEndSepPunct{\mcitedefaultmidpunct}
{\mcitedefaultendpunct}{\mcitedefaultseppunct}\relax
\EndOfBibitem
\bibitem[Yanagishita \latin{et~al.}(2009)Yanagishita, Nishio, and
  Masuda]{Yanagishita}
Yanagishita,~T.; Nishio,~K.; Masuda,~H. Anti-reflection structures on lenses by
  nanoimprinting using ordered anodic porous alumina. \emph{Appl. Phys.
  Express} \textbf{2009}, \emph{2}, 022001\relax
\mciteBstWouldAddEndPuncttrue
\mciteSetBstMidEndSepPunct{\mcitedefaultmidpunct}
{\mcitedefaultendpunct}{\mcitedefaultseppunct}\relax
\EndOfBibitem
\bibitem[Wilt \latin{et~al.}(2018)Wilt, Goul, Acharya, Sakidja, and Wu]{Wilt}
Wilt,~J.; Goul,~R.; Acharya,~J.; Sakidja,~R.; Wu,~J.~Z. In situ atomic layer
  deposition and electron tunneling characterization of monolayer Al$_2$O$_3$
  on Fe for magnetic tunnel junctions. \emph{AIP Adv.} \textbf{2018}, \emph{8},
  125218\relax
\mciteBstWouldAddEndPuncttrue
\mciteSetBstMidEndSepPunct{\mcitedefaultmidpunct}
{\mcitedefaultendpunct}{\mcitedefaultseppunct}\relax
\EndOfBibitem
\bibitem[Meng \latin{et~al.}(2014)Meng, Liu, Zhang, Ren, and Jia]{Meng}
Meng,~F.~C.; Liu,~C.; Zhang,~X.~L.; Ren,~H.; Jia,~T. Synthesis and
  Characterization of $\gamma$-Al$_2$O$_3$ Nanowires by a Surfactant Assisted
  Precipitation Method. \emph{Key Eng. Mater.} \textbf{2014}, \emph{602-603},
  93 -- 96\relax
\mciteBstWouldAddEndPuncttrue
\mciteSetBstMidEndSepPunct{\mcitedefaultmidpunct}
{\mcitedefaultendpunct}{\mcitedefaultseppunct}\relax
\EndOfBibitem
\bibitem[Li \latin{et~al.}(2017)Li, Kong, Huang, and Li]{Li2}
Li,~Z.; Kong,~L.; Huang,~S.; Li,~L. Highly luminescent and ultrastable
  CsPbBr$_3$ perovskite quantum dots incorporated into a silica/alumina
  monolith. \emph{Angew. Chem} \textbf{2017}, \emph{129}, 8246--8250\relax
\mciteBstWouldAddEndPuncttrue
\mciteSetBstMidEndSepPunct{\mcitedefaultmidpunct}
{\mcitedefaultendpunct}{\mcitedefaultseppunct}\relax
\EndOfBibitem
\bibitem[Napari \latin{et~al.}(2021)Napari, Huq, Meeth, Heikkilä, Niang, Wang,
  Iivonen, Wang, Leskelä, Ritala, Flewitt, Hoye, and
  MacManus-Driscoll]{Napari}
Napari,~M.; Huq,~T.; Meeth,~D.; Heikkilä,~M.; Niang,~K.; Wang,~H.;
  Iivonen,~T.; Wang,~H.; Leskelä,~M.; Ritala,~M.; Flewitt,~A.; Hoye,~R.;
  MacManus-Driscoll,~J. Role of ALD Al 2 O 3 Surface Passivation on the
  Performance of p-Type Cu$_2$O Thin Film Transistors. \emph{ACS Appl. Mater.
  Interfaces} \textbf{2021}, \emph{13}, 4156--4164\relax
\mciteBstWouldAddEndPuncttrue
\mciteSetBstMidEndSepPunct{\mcitedefaultmidpunct}
{\mcitedefaultendpunct}{\mcitedefaultseppunct}\relax
\EndOfBibitem
\bibitem[Song \latin{et~al.}(2016)Song, Yang, Chai, Callsen, Zhou, Yang, Zhang,
  PAN, Chi, Feng, and Wang]{Song}
Song,~T.; Yang,~M.; Chai,~J.; Callsen,~M.; Zhou,~J.; Yang,~T.; Zhang,~Z.;
  PAN,~J.; Chi,~D.; Feng,~Y.~P.; Wang,~S. The stability of aluminium oxide
  monolayer and its interface with two-dimensional materials. \emph{Sci. Rep.}
  \textbf{2016}, \emph{6}, 29221\relax
\mciteBstWouldAddEndPuncttrue
\mciteSetBstMidEndSepPunct{\mcitedefaultmidpunct}
{\mcitedefaultendpunct}{\mcitedefaultseppunct}\relax
\EndOfBibitem
\bibitem[Cheng \latin{et~al.}(2015)Cheng, Tomczyk, Lu, Veazey, Huang, Irvin,
  Ryu, Lee, Eom, Hellberg, and Levy]{Cheng}
Cheng,~G.; Tomczyk,~M.; Lu,~S.; Veazey,~J.; Huang,~M.; Irvin,~P.; Ryu,~S.;
  Lee,~H.; Eom,~C.-B.; Hellberg,~C.; Levy,~J. Electron pairing without
  superconductivity. \emph{Nature (London)} \textbf{2015}, \emph{521},
  196--9\relax
\mciteBstWouldAddEndPuncttrue
\mciteSetBstMidEndSepPunct{\mcitedefaultmidpunct}
{\mcitedefaultendpunct}{\mcitedefaultseppunct}\relax
\EndOfBibitem
\bibitem[Ghosh \latin{et~al.}(2020)Ghosh, Kumar, Roy, Nguyen, Ahuja, Adil,
  Chatti, Kar, MacFarlane, and Mitra]{Ghosh}
Ghosh,~A.; Kumar,~A.; Roy,~A.; Nguyen,~C.; Ahuja,~A.; Adil,~M.; Chatti,~M.;
  Kar,~M.; MacFarlane,~D.~R.; Mitra,~S. Ultrathin Lithium Aluminate
  Nanoflake-Inlaid Sulfur as a Cathode Material for Lithium--Sulfur Batteries
  with High Areal Capacity. \emph{ACS Appl. Energy Mater.} \textbf{2020},
  \emph{3}, 5637--5645\relax
\mciteBstWouldAddEndPuncttrue
\mciteSetBstMidEndSepPunct{\mcitedefaultmidpunct}
{\mcitedefaultendpunct}{\mcitedefaultseppunct}\relax
\EndOfBibitem
\bibitem[Li \latin{et~al.}(2015)Li, Budai, Liu, Chen, Howe, Sun, Tischler,
  Meltzer, and Pan]{Li}
Li,~X.; Budai,~J.~D.; Liu,~F.; Chen,~Y.-S.; Howe,~J.~Y.; Sun,~C.;
  Tischler,~J.~Z.; Meltzer,~R.~S.; Pan,~Z. Crystal structures and optical
  properties of new quaternary strontium europium aluminate luminescent
  nanoribbons. \emph{J. Mater. Chem. C} \textbf{2015}, \emph{3}, 778--788\relax
\mciteBstWouldAddEndPuncttrue
\mciteSetBstMidEndSepPunct{\mcitedefaultmidpunct}
{\mcitedefaultendpunct}{\mcitedefaultseppunct}\relax
\EndOfBibitem
\bibitem[Demirci \latin{et~al.}(2017)Demirci, Avazl{\i}, Durgun, and
  Cahangirov]{Demirci}
Demirci,~S.; Avazl{\i},~N.; Durgun,~E.; Cahangirov,~S. Structural and
  electronic properties of monolayer group III monochalcogenides. \emph{Phys.
  Rev. B} \textbf{2017}, \emph{95}, 115409\relax
\mciteBstWouldAddEndPuncttrue
\mciteSetBstMidEndSepPunct{\mcitedefaultmidpunct}
{\mcitedefaultendpunct}{\mcitedefaultseppunct}\relax
\EndOfBibitem
\bibitem[Novoselov \latin{et~al.}(2004)Novoselov, Geim, Morozov, Jiang, Zhang,
  Dubonos, Grigorieva, and Firsov]{novoselov2004electric}
Novoselov,~K.~S.; Geim,~A.~K.; Morozov,~S.~V.; Jiang,~D.-e.; Zhang,~Y.;
  Dubonos,~S.~V.; Grigorieva,~I.~V.; Firsov,~A.~A. Electric field effect in
  atomically thin carbon films. \emph{Science} \textbf{2004}, \emph{306},
  666--669\relax
\mciteBstWouldAddEndPuncttrue
\mciteSetBstMidEndSepPunct{\mcitedefaultmidpunct}
{\mcitedefaultendpunct}{\mcitedefaultseppunct}\relax
\EndOfBibitem
\bibitem[Geim(2009)]{geim2009graphene}
Geim,~A.~K. Graphene: status and prospects. \emph{Science} \textbf{2009},
  \emph{324}, 1530--1534\relax
\mciteBstWouldAddEndPuncttrue
\mciteSetBstMidEndSepPunct{\mcitedefaultmidpunct}
{\mcitedefaultendpunct}{\mcitedefaultseppunct}\relax
\EndOfBibitem
\bibitem[Neto \latin{et~al.}(2009)Neto, Guinea, Peres, Novoselov, and
  Geim]{Neto}
Neto,~A.~C.; Guinea,~F.; Peres,~N.~M.; Novoselov,~K.~S.; Geim,~A.~K. The
  electronic properties of graphene. \emph{Rev. Mod. Phys} \textbf{2009},
  \emph{81}, 109\relax
\mciteBstWouldAddEndPuncttrue
\mciteSetBstMidEndSepPunct{\mcitedefaultmidpunct}
{\mcitedefaultendpunct}{\mcitedefaultseppunct}\relax
\EndOfBibitem
\bibitem[Lee \latin{et~al.}(2008)Lee, Wei, Kysar, and Hone]{lee2008measurement}
Lee,~C.; Wei,~X.; Kysar,~J.~W.; Hone,~J. Measurement of the elastic properties
  and intrinsic strength of monolayer graphene. \emph{Science} \textbf{2008},
  \emph{321}, 385--388\relax
\mciteBstWouldAddEndPuncttrue
\mciteSetBstMidEndSepPunct{\mcitedefaultmidpunct}
{\mcitedefaultendpunct}{\mcitedefaultseppunct}\relax
\EndOfBibitem
\bibitem[Li and Chen(2016)Li, and Chen]{Li3}
Li,~L.~H.; Chen,~Y. Atomically thin boron nitride: unique properties and
  applications. \emph{Adv. Funct. Mater.} \textbf{2016}, \emph{26},
  2594--2608\relax
\mciteBstWouldAddEndPuncttrue
\mciteSetBstMidEndSepPunct{\mcitedefaultmidpunct}
{\mcitedefaultendpunct}{\mcitedefaultseppunct}\relax
\EndOfBibitem
\bibitem[Mak \latin{et~al.}(2010)Mak, Lee, Hone, Shan, and Heinz]{Mak}
Mak,~K.~F.; Lee,~C.; Hone,~J.; Shan,~J.; Heinz,~T.~F. Atomically thin MoS$_2$:
  a new direct-gap semiconductor. \emph{Phys. Rev. Lett.} \textbf{2010},
  \emph{105}, 136805\relax
\mciteBstWouldAddEndPuncttrue
\mciteSetBstMidEndSepPunct{\mcitedefaultmidpunct}
{\mcitedefaultendpunct}{\mcitedefaultseppunct}\relax
\EndOfBibitem
\bibitem[Wang \latin{et~al.}(2012)Wang, Kalantar-Zadeh, Kis, Coleman, and
  Strano]{wang2012electronics}
Wang,~Q.~H.; Kalantar-Zadeh,~K.; Kis,~A.; Coleman,~J.~N.; Strano,~M.~S.
  Electronics and optoelectronics of two-dimensional transition metal
  dichalcogenides. \emph{Nat. Nanotechnol.} \textbf{2012}, \emph{7},
  699--712\relax
\mciteBstWouldAddEndPuncttrue
\mciteSetBstMidEndSepPunct{\mcitedefaultmidpunct}
{\mcitedefaultendpunct}{\mcitedefaultseppunct}\relax
\EndOfBibitem
\bibitem[Tongay \latin{et~al.}(2014)Tongay, Sahin, Ko, Luce, Fan, Liu, Zhou,
  Huang, Ho, Yan, \latin{et~al.} others]{tongay2014monolayer}
others,, \latin{et~al.}  Monolayer behaviour in bulk ReS$_2$ due to electronic
  and vibrational decoupling. \emph{Nat. Commun.} \textbf{2014}, \emph{5},
  1--6\relax
\mciteBstWouldAddEndPuncttrue
\mciteSetBstMidEndSepPunct{\mcitedefaultmidpunct}
{\mcitedefaultendpunct}{\mcitedefaultseppunct}\relax
\EndOfBibitem
\bibitem[Cahangirov \latin{et~al.}(2009)Cahangirov, Topsakal, Akt{\"u}rk,
  {\c{S}}ahin, and Ciraci]{silicene2009}
Cahangirov,~S.; Topsakal,~M.; Akt{\"u}rk,~E.; {\c{S}}ahin,~H.; Ciraci,~S.
  Two-and one-dimensional honeycomb structures of silicon and germanium.
  \emph{Phys. Rev. Lett.} \textbf{2009}, \emph{102}, 236804\relax
\mciteBstWouldAddEndPuncttrue
\mciteSetBstMidEndSepPunct{\mcitedefaultmidpunct}
{\mcitedefaultendpunct}{\mcitedefaultseppunct}\relax
\EndOfBibitem
\bibitem[Deng \latin{et~al.}(2018)Deng, Yu, Song, Zhang, Wang, Sun, Yi, Wu, Wu,
  Zhu, and et~al.]{Deng}
Deng,~Y.; Yu,~Y.; Song,~Y.; Zhang,~J.; Wang,~N.~Z.; Sun,~Z.; Yi,~Y.; Wu,~Y.~Z.;
  Wu,~S.; Zhu,~J.; et~al., Gate-tunable room-temperature ferromagnetism in
  two-dimensional Fe$_3$GeTe$_2$. \emph{Nature (London)} \textbf{2018},
  \emph{563}, 94–99\relax
\mciteBstWouldAddEndPuncttrue
\mciteSetBstMidEndSepPunct{\mcitedefaultmidpunct}
{\mcitedefaultendpunct}{\mcitedefaultseppunct}\relax
\EndOfBibitem
\bibitem[Li \latin{et~al.}(2021)Li, Ren, Wan, Liu, and Ge]{Li4}
Li,~F.; Ren,~Y.; Wan,~W.; Liu,~Y.; Ge,~Y. Strain tunable intrinsic
  ferromagnetic in 2D square CrBr$_2$. \emph{AIP Adv.} \textbf{2021},
  \emph{11}, 115220\relax
\mciteBstWouldAddEndPuncttrue
\mciteSetBstMidEndSepPunct{\mcitedefaultmidpunct}
{\mcitedefaultendpunct}{\mcitedefaultseppunct}\relax
\EndOfBibitem
\bibitem[Xuhan \latin{et~al.}(2020)Xuhan, Brzostowski, Durajski, Liu, Xiang,
  Jiang, Wang, Chen, Li, Zhong, Drzewiński, Jarosik, Szczęśniak, Lai, Guo,
  and Zhong]{zhou2020atomically}
Xuhan,~Z. \latin{et~al.}  Atomically Thin 1T-FeCl$_2$ Grown by Molecular-Beam
  Epitaxy. \emph{J. Phys. Chem. C} \textbf{2020}, \emph{124}, 9416--9423\relax
\mciteBstWouldAddEndPuncttrue
\mciteSetBstMidEndSepPunct{\mcitedefaultmidpunct}
{\mcitedefaultendpunct}{\mcitedefaultseppunct}\relax
\EndOfBibitem
\bibitem[Ceyhan \latin{et~al.}(2021)Ceyhan, Yagmurcukardes, Peeters, and
  Sahin]{ceyhan2021}
Ceyhan,~E.; Yagmurcukardes,~M.; Peeters,~F.; Sahin,~H. Electronic and magnetic
  properties of single-layer FeCl$_2$ with defects. \emph{Phys. Rev. B}
  \textbf{2021}, \emph{103}, 014106\relax
\mciteBstWouldAddEndPuncttrue
\mciteSetBstMidEndSepPunct{\mcitedefaultmidpunct}
{\mcitedefaultendpunct}{\mcitedefaultseppunct}\relax
\EndOfBibitem
\bibitem[Grimme \latin{et~al.}(2010)Grimme, Antony, Ehrlich, and Krieg]{Grimme}
Grimme,~S.; Antony,~J.; Ehrlich,~S.; Krieg,~H. A consistent and accurate ab
  initio parametrization of density functional dispersion correction (DFT-D)
  for the 94 elements H-Pu. \emph{J. Chem. Phys.} \textbf{2010}, \emph{132},
  154104\relax
\mciteBstWouldAddEndPuncttrue
\mciteSetBstMidEndSepPunct{\mcitedefaultmidpunct}
{\mcitedefaultendpunct}{\mcitedefaultseppunct}\relax
\EndOfBibitem
\bibitem[Kresse and Furthm{\"u}ller(1996)Kresse, and Furthm{\"u}ller]{Kresse}
Kresse,~G.; Furthm{\"u}ller,~J. Efficient iterative schemes for ab initio
  total-energy calculations using a plane-wave basis set. \emph{Phys. Rev. B}
  \textbf{1996}, \emph{54}, 11169\relax
\mciteBstWouldAddEndPuncttrue
\mciteSetBstMidEndSepPunct{\mcitedefaultmidpunct}
{\mcitedefaultendpunct}{\mcitedefaultseppunct}\relax
\EndOfBibitem
\bibitem[Kresse and Hafner(1993)Kresse, and Hafner]{Kresse2}
Kresse,~G.; Hafner,~J. Ab initio molecular dynamics for liquid metals.
  \emph{Phys. Rev. B} \textbf{1993}, \emph{47}, 558\relax
\mciteBstWouldAddEndPuncttrue
\mciteSetBstMidEndSepPunct{\mcitedefaultmidpunct}
{\mcitedefaultendpunct}{\mcitedefaultseppunct}\relax
\EndOfBibitem
\bibitem[Perdew \latin{et~al.}(1996)Perdew, Burke, and Ernzerhof]{Perdew}
Perdew,~J.~P.; Burke,~K.; Ernzerhof,~M. Generalized gradient approximation made
  simple. \emph{Phys. Rev. Lett.} \textbf{1996}, \emph{77}, 3865\relax
\mciteBstWouldAddEndPuncttrue
\mciteSetBstMidEndSepPunct{\mcitedefaultmidpunct}
{\mcitedefaultendpunct}{\mcitedefaultseppunct}\relax
\EndOfBibitem
\bibitem[Bl{\"o}chl(1994)]{Blochl}
Bl{\"o}chl,~P.~E. Projector augmented-wave method. \emph{Phys. Rev. B}
  \textbf{1994}, \emph{50}, 17953\relax
\mciteBstWouldAddEndPuncttrue
\mciteSetBstMidEndSepPunct{\mcitedefaultmidpunct}
{\mcitedefaultendpunct}{\mcitedefaultseppunct}\relax
\EndOfBibitem
\bibitem[Henkelman \latin{et~al.}(2006)Henkelman, Arnaldsson, and
  J{\'o}nsson]{Henkelman}
Henkelman,~G.; Arnaldsson,~A.; J{\'o}nsson,~H. A fast and robust algorithm for
  Bader decomposition of charge density. \emph{Comput. Mater. Sci.}
  \textbf{2006}, \emph{36}, 354--360\relax
\mciteBstWouldAddEndPuncttrue
\mciteSetBstMidEndSepPunct{\mcitedefaultmidpunct}
{\mcitedefaultendpunct}{\mcitedefaultseppunct}\relax
\EndOfBibitem
\bibitem[Togo and Tanaka(2015)Togo, and Tanaka]{Togo}
Togo,~A.; Tanaka,~I. First principles phonon calculations in materials science.
  \emph{Scr. Mater.} \textbf{2015}, \emph{108}, 1--5\relax
\mciteBstWouldAddEndPuncttrue
\mciteSetBstMidEndSepPunct{\mcitedefaultmidpunct}
{\mcitedefaultendpunct}{\mcitedefaultseppunct}\relax
\EndOfBibitem
\bibitem[Lanzillo \latin{et~al.}(2015)Lanzillo, Simbeck, and Nayak]{Lanzillo}
Lanzillo,~N.~A.; Simbeck,~A.~J.; Nayak,~S.~K. Strain engineering the work
  function in monolayer metal dichalcogenides. \emph{J. Phys. Condens. Matter}
  \textbf{2015}, \emph{27}, 175501\relax
\mciteBstWouldAddEndPuncttrue
\mciteSetBstMidEndSepPunct{\mcitedefaultmidpunct}
{\mcitedefaultendpunct}{\mcitedefaultseppunct}\relax
\EndOfBibitem
\bibitem[Pierucci \latin{et~al.}(2018)Pierucci, Zribi, Henck, Chaste, Silly,
  Bertran, Fevre, Gil, Summerfield, Beton, Novikov, Cassabois, Rault, and
  Ouerghi]{Pierucci}
Pierucci,~D.; Zribi,~J.; Henck,~H.; Chaste,~J.; Silly,~M.; Bertran,~F.;
  Fevre,~P.; Gil,~B.; Summerfield,~A.; Beton,~P.; Novikov,~S.; Cassabois,~G.;
  Rault,~J.; Ouerghi,~A. Van der Waals epitaxy of two-dimensional single-layer
  h-BN on graphite by molecular beam epitaxy: Electronic properties and band
  structure. \emph{Appl. Phys. Lett.} \textbf{2018}, \emph{112}, 253102\relax
\mciteBstWouldAddEndPuncttrue
\mciteSetBstMidEndSepPunct{\mcitedefaultmidpunct}
{\mcitedefaultendpunct}{\mcitedefaultseppunct}\relax
\EndOfBibitem
\bibitem[Bivour \latin{et~al.}(2015)Bivour, Temmler, Steinkemper, and
  Hermle]{Bivour}
Bivour,~M.; Temmler,~J.; Steinkemper,~H.; Hermle,~M. Molybdenum and tungsten
  oxide: High work function wide band gap contact materials for hole selective
  contacts of silicon solar cells. \emph{Sol. Energy Mater. Sol. Cells}
  \textbf{2015}, \emph{142}, 34--41\relax
\mciteBstWouldAddEndPuncttrue
\mciteSetBstMidEndSepPunct{\mcitedefaultmidpunct}
{\mcitedefaultendpunct}{\mcitedefaultseppunct}\relax
\EndOfBibitem
\bibitem[Schulz \latin{et~al.}(2016)Schulz, Tiepelt, Christians, Levine, Edri,
  Sanehira, Hodes, Cahen, and Kahn]{Schulz}
Schulz,~P.; Tiepelt,~J.~O.; Christians,~J.~A.; Levine,~I.; Edri,~E.;
  Sanehira,~E.~M.; Hodes,~G.; Cahen,~D.; Kahn,~A. High-work-function molybdenum
  oxide hole extraction contacts in hybrid organic--inorganic perovskite solar
  cells. \emph{ACS Appl. Mater. Interfaces} \textbf{2016}, \emph{8},
  31491--31499\relax
\mciteBstWouldAddEndPuncttrue
\mciteSetBstMidEndSepPunct{\mcitedefaultmidpunct}
{\mcitedefaultendpunct}{\mcitedefaultseppunct}\relax
\EndOfBibitem
\bibitem[Ma \latin{et~al.}(2018)Ma, Kou, Huang, Dai, and Heine]{ma2018two}
Ma,~Y.; Kou,~L.; Huang,~B.; Dai,~Y.; Heine,~T. Two-dimensional ferroelastic
  topological insulators in single-layer Janus transition metal dichalcogenides
  M SSe (M= Mo, W). \emph{Phys. Rev. B} \textbf{2018}, \emph{98}, 085420\relax
\mciteBstWouldAddEndPuncttrue
\mciteSetBstMidEndSepPunct{\mcitedefaultmidpunct}
{\mcitedefaultendpunct}{\mcitedefaultseppunct}\relax
\EndOfBibitem
\bibitem[{\c{S}}ahin \latin{et~al.}(2009){\c{S}}ahin, Cahangirov, Topsakal,
  Bekaroglu, Akturk, Senger, and Ciraci]{sahin35}
{\c{S}}ahin,~H.; Cahangirov,~S.; Topsakal,~M.; Bekaroglu,~E.; Akturk,~E.;
  Senger,~R.~T.; Ciraci,~S. Monolayer honeycomb structures of group-IV elements
  and III-V binary compounds: First-principles calculations. \emph{Phys. Rev.
  B} \textbf{2009}, \emph{80}, 155453\relax
\mciteBstWouldAddEndPuncttrue
\mciteSetBstMidEndSepPunct{\mcitedefaultmidpunct}
{\mcitedefaultendpunct}{\mcitedefaultseppunct}\relax
\EndOfBibitem
\bibitem[Gercek(2007)]{Gercek}
Gercek,~H. Poisson's ratio values for rocks. \emph{Int. J. Rock Mech. Min.
  Sci.} \textbf{2007}, \emph{44}, 1--13\relax
\mciteBstWouldAddEndPuncttrue
\mciteSetBstMidEndSepPunct{\mcitedefaultmidpunct}
{\mcitedefaultendpunct}{\mcitedefaultseppunct}\relax
\EndOfBibitem
\bibitem[Yagmurcukardes \latin{et~al.}(2016)Yagmurcukardes, Senger, Peeters,
  and Sahin]{Yagmurcukardes}
Yagmurcukardes,~M.; Senger,~R.~T.; Peeters,~F.~M.; Sahin,~H. Mechanical
  properties of monolayer GaS and GaSe crystals. \emph{Phys. Rev. B}
  \textbf{2016}, \emph{94}, 245407\relax
\mciteBstWouldAddEndPuncttrue
\mciteSetBstMidEndSepPunct{\mcitedefaultmidpunct}
{\mcitedefaultendpunct}{\mcitedefaultseppunct}\relax
\EndOfBibitem
\bibitem[Ionescu and Riel(2011)Ionescu, and Riel]{Ionescu}
Ionescu,~A.~M.; Riel,~H. Tunnel field-effect transistors as energy-efficient
  electronic switches. \emph{Nature (London)} \textbf{2011}, \emph{479},
  329--337\relax
\mciteBstWouldAddEndPuncttrue
\mciteSetBstMidEndSepPunct{\mcitedefaultmidpunct}
{\mcitedefaultendpunct}{\mcitedefaultseppunct}\relax
\EndOfBibitem
\bibitem[Lu and Seabaugh(2014)Lu, and Seabaugh]{Lu}
Lu,~H.; Seabaugh,~A. Tunnel field-effect transistors: State-of-the-art.
  \emph{IEEE J. Electron Devices Soc.} \textbf{2014}, \emph{2}, 44--49\relax
\mciteBstWouldAddEndPuncttrue
\mciteSetBstMidEndSepPunct{\mcitedefaultmidpunct}
{\mcitedefaultendpunct}{\mcitedefaultseppunct}\relax
\EndOfBibitem
\bibitem[Xia \latin{et~al.}(2018)Xia, Du, Li, Li, Zhao, Wang, and Li]{BTT}
Xia,~C.; Du,~J.; Li,~M.; Li,~X.; Zhao,~X.; Wang,~T.; Li,~J. Effects of Electric
  Field on the Electronic Structures of Broken-gap Phosphorene/SnX$_2$ (X= S,
  Se) van der Waals Heterojunctions. \emph{Phys. Rev. Appl.} \textbf{2018},
  \emph{10}, 054064\relax
\mciteBstWouldAddEndPuncttrue
\mciteSetBstMidEndSepPunct{\mcitedefaultmidpunct}
{\mcitedefaultendpunct}{\mcitedefaultseppunct}\relax
\EndOfBibitem
\end{mcitethebibliography}
\bibliographystyle{achemso}

\end{document}